# Informative Sensing

Hyun Sung Chang[1]     Yair Weiss[2]     William T. Freeman[1]

[1]Computer Science and Artificial Intelligence Laboratory, MIT, USA

[2]School of Computer Science and Engineering, The Hebrew University of Jerusalem, Israel

## Abstract

Compressed sensing is a recent set of mathematical results showing that sparse signals can be exactly reconstructed from a small number of linear measurements. Interestingly, for ideal sparse signals with no measurement noise, random measurements allow perfect reconstruction while measurements based on principal component analysis (PCA) or independent component analysis (ICA) do not. At the same time, for other signal and noise distributions, PCA and ICA can significantly outperform random projections in terms of enabling reconstruction from a small number of measurements. In this paper we ask: given the distribution of signals we wish to measure, what are the optimal set of linear projections for compressed sensing? We consider the problem of finding a small number of linear projections that are maximally informative about the signal. Formally, we use the InfoMax criterion and seek to maximize the mutual information between the signal, $x$, and the (possibly noisy) projection $y = Wx$. We show that in general the optimal projections are not the principal components of the data nor random projections, but rather a seemingly novel set of projections that capture what is still uncertain about the signal, given the knowledge of distribution. We present analytic solutions for certain special cases including natural images. In particular, for natural images, the near-optimal projections are bandwise random, i.e., incoherent to the sparse bases at a particular frequency band but with more weights on the low-frequencies, which has a physical relation to the multi-resolution representation of images.

## I. Introduction

Compressed sensing [1], [2] is a set of recent mathematical results on a classic question: given a signal $x \in \mathbb{R}^d$ and a set of $p$ linear measurements $y = Wx \in \mathbb{R}^p$, how many measurements are required to enable reconstruction of $x$? Obviously, if we knew nothing at all about $x$, i.e. $x$ can be any $d$ dimensional vector, we would need $d$ measurements. Alternatively, if we know our signal $x$ lies in a low-dimensional linear subspace, say of dimension $k$, then $k$ measurements are sufficient. But what if we know that $x$ lies in a low-dimensional *nonlinear* manifold? Can we still get away with fewer than $d$ measurements?

To motivate this question, consider the space of natural images. An image with $d$ pixels can be thought of as a vector in $\mathbb{R}^d$, but natural images occupy a tiny fraction of the set of all signals in this space. If there was a way to exploit this fact, we could build cameras with a small number of sensors that would still enable us perfect, high resolution, reconstructions for natural images.

The basic mathematical results in compressed sensing deal with signals that are $k$ sparse. These are signals that can be represented with a small number, $k$ of active (non-zero) basis elements. For such signals, it was shown in [1], [2], that $ck \log d$ *generic* linear measurements are sufficient to recover the signal exactly (with $c$ a constant). Furthermore, the recovery can be done by a simple convex optimization or by a greedy optimization procedure [3].

These results have generated a tremendous amount of excitement in both the theoretical and practical communities. On the theoretical side, the performance of compressed sensing with random projections has been analyzed when the signals are not exactly $k$ sparse, but rather *compressible* (i.e. can be well approximated with a small number of active basis elements) [1], [2] as well as when the measurements are contaminated with noise [4]–[6]. On the practical side, applications of compressed sensing have been explored in building "single-pixel" cameras [7], medical imaging [8], [9] and geophysical data analysis [10].

Perhaps the most surprising result in compressed sensing is that perfect recovery is possible with *random projections*. This is surprising given the large amount of literature in machine learning and statistics devoted to finding projections that are optimal in some sense (e.g. [11]). In fact, as we see in this paper, for white sparse signals, *random measurements* significantly outperform measurements based on principal component analysis (PCA) or independent component analysis (ICA). At the same time, for other signal and noise distributions, PCA and ICA can significantly outperform random projections.

In this paper we ask: given a distribution or statistics of the signals we wish to measure, what are the optimal set of linear projections for compressed sensing? We show that the optimal projections are in general not the principal components nor the independent components of the data, but rather a seemingly novel set of projections that capture what is still uncertain about the signal, given the signal distribution. We present analytic solutions for various special cases, including natural images, and demonstrate, by experiments, that the projections onto the uncertain components may far outperform random projections.

## II. INFORMATIVE SENSING

In [12], Linsker suggested the InfoMax principle for the design of a linear sensory system. According to this principle, the goal of the sensory system is to maximize the mutual information between the sensors and the world (see also [13]–[15]). In this paper, we are interested in *undercomplete* InfoMax where the number of sensors is less than the dimensionality of the input. Given that dimensionality reduction throws away some information about the input, does it still make sense to maximize mutual information in the undercomplete case?

Consider an example, shown in Fig. 1, where a 1-dimensional measurement is taken on a 2-dimensional signal distributed in a mixture of four Gaussians. The black line in each figure denotes a projection vector, and the red line shows the Bayes least-squares (BLS) estimate of signals given the projection. When a random projection (left) or a single PCA projection (middle) are used, the BLS estimate is quite far from the original data, indicating that a lot of information has been lost. But the InfoMax projection (shown on the right) provides much more information, making the BLS decoding significantly better.

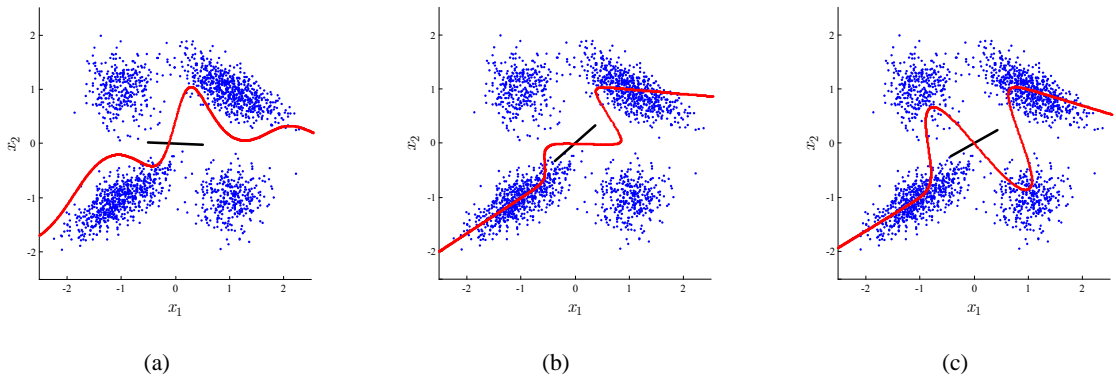

|     |     |     |
| --- | --- | --- |
| (a) | (b) | (c) |

Fig. 1. Different kinds of one-dimensional projection schemes and their reconstruction results for a two-dimensional Gaussian mixture source. (a) Random, (b) PCA, (c) InfoMax projections. Blue points: data samples. Black line: projection vector. Red curve: reconstruction based on the BLS estimate, $\hat{x} = E[x|y]$.

The key to this result lies in the nonlinearity of the decoding scheme. It is well known that projections based on PCA give optimal reconstruction, in terms of mean squared error (MSE), if the decoding is restricted to be linear. But, if the decoding is allowed to be nonlinear, the optimal projection may significantly differ from the PCA projection. In this regard, compressed sensing of sparse signals [1], [2] is a spectacular demonstration of nonlinear decoding from a small number of linear projections.

Let $x$ and $y$ be the sensor input (original signal) and sensor output (measurements) related by $y = Wx + \eta$, where $W$ is a $p \times d$ matrix ($p < d$) and $\eta$ represents the sensor noise assumed to be Gaussian, $\eta \sim G(0, \sigma^2 I)$. Our problem may be formally defined as

$$W^* = \arg\max_{W} I(x; Wx + \eta). \tag{1}$$

Without constraint on $W$, the mutual information can be made arbitrarily large, simply by scaling $W$. To preclude such a trivial manipulation, we restrict $W$ to satisfy the orthonormality condition, i.e., $WW^T = I$.[1]

Note that (1) looks quite similar to the definition of channel capacity. Considering the sensing process as a channel, we may call $I(x; Wx + \eta)$ *sensor capacity*, which measures how informative the sensor is. However, as noted by Linsker, there is a crucial difference between the channel capacity problem and the InfoMax problem: In (1), the source signal has a certain probability distribution, while the optimal choice of channel $W$ is actually being sought for, which is exactly the reverse case of the channel capacity problem.

A simple alternative to (1) is to use $h(y)$, which characterizes the information content of the measurements [6], instead of $I(x; y)$ because $I(x; y) = h(y) - h(y|x)$ and $h(y|x)$ is merely the entropy of the noise $\eta$, which is invariant to $W$. Besides its simplicity, this objective function has another desirable property that it is well defined even without noise, unlike $I(x; Wx)$ which diverges to infinity. In Appendix A, we will further discuss the validity of $h(Wx)$ with $WW^T = I$ as an objective function for the noiseless condition.

The values of $h(Wx)$ for the sensing schemes in Fig. 1 are numerically evaluated and compared in Table I. Indeed, the scheme with highest entropy corresponds to the best reconstruction. Although this needs not always be the case, InfoMax and reconstruction are closely related. This can be seen by considering the cost function suggested in [16]: $L(W) = \prod_n \Pr(x_n|y_n; W)$ where $x_n$ are samples drawn from $\Pr(x)$ and $y_n = Wx_n$. It is easy to see that $\ln L(W) \to -h(x|y)$ so that maximizing $h(y)$ is equivalent to maximizing the probability of correct reconstruction of a sample given its projection.

[1]In certain applications, it makes more sense to limit the total available power and replace the orthonormality constraint with $\mathrm{tr}(WW^T) \le P$ for a fixed budget $P$. For the noiseless case, the two constraints can be shown to result in the same optimal set of projections.

 

TABLE I

Performance Comparison among Three Projection Schemes in Example of Fig. 1, in terms of Entropy $h(Wx)$ and Mean Squared Error (MSE).

|        | a     | b     | c     |
|--------|-------|-------|-------|
| $h(y)$ | 1.189 | 1.345 | 1.449 |
| MSE    | 0.931 | 0.726 | 0.476 |

Minimizing uncertainty has also been proposed recently in the context of sequential design of compressed sensing by [6], [17]. In this context, the projections are chosen sequentially where each projection mimimizes the remaining uncertainty about the signal given the results of the previous projection.

## III. Analysis

The optimal projections maximizing $h(y)$ vary according to the prior distribution of the source signal $x$. Multi-variate Gaussian is a special kind of signal whose optimal projection can be found analytically, but, in most cases, the optimal projection is hard to find in a closed-form because of the complicated nature of differential entropy.

For a $p$-dimensional random vector $y$ whose covariance is $\Sigma_y$, its entropy $h(y)$ can be decomposed into

$$h(y) = h(\widetilde{y}) + \frac{1}{2}\ln\det(\Sigma_y) \tag{2}$$

where $\widetilde{y}$ is a whitened version of $y$, e.g., $\widetilde{y} = \Sigma_y^{-\frac{1}{2}} y$. Note that $h(\widetilde{y})$ is covariance-free, depending only on the shape of the probability distribution of $\widetilde{y}$. It is well known that $h(\widetilde{y})$ is maximized to $\frac{p}{2}\ln(2\pi e)$ if and only if $\widetilde{y}$ is jointly Gaussian. On the other hand, the second term, $\frac{1}{2}\ln\det(\Sigma_y)$, depends only on $\Sigma_y$, the covariance of $y$. So, we will call $h(\widetilde{y})$ the *shape term* and $\frac{1}{2}\ln\det(\Sigma_y)$ the *variance term*. Overall, an entropy is a sum of the shape term and the variance term.

In the following, we present some analytical results for undercomplete InfoMax, based on the entropy decomposition, and make comparisons between two popular projection schemes (PCA and random projections) for various special cases, which provides us better understanding on the desirable behaviors for the informative sensing.

## A. White Signals

*Observation 1:* For white data, the projected signal must be as Gaussian as possible to be most informative.

*Proof:* For white data, the variance term goes away and only the shape term remains. Therefore, the InfoMax is really achieved by the projection which can maximize the shape term. ∎

*Observation 2:* For white data $x$ with zero-mean and finite variances, if the distribution of $\|x\| / \sqrt{d}$ is degenerated to a constant, $p$ random projections, for $p < O(\sqrt{d})$, are asymptotically optimal if $d$ goes to infinity.

*Proof:* In [18], Dasgupta *et al.* have shown that almost all $p$ linear projections behave like a scale-mixture of zero-mean Gaussians with variances that have a profile that is the same as the distribution of $\|x\| / \sqrt{d}$. If $\|x\| / \sqrt{d}$ goes to a Dirac's delta function, the mixture will collapse to a single Gaussian. Specifically, in this case, the main theorem of [18] reads approximately like the following: For any ball $B \in \mathbb{R}^p$ and for almost all $W$,

$$\sup_{B \in \mathbb{R}^p} \left| \Pr(B; W) - \overline{\Pr}(B) \right| \leq O(p^2/d)^{1/4} \tag{3}$$

where $\sup$ represents the supremum and where $\Pr(\cdot; W)$ and $\overline{\Pr}(\cdot)$ denote the probability with respect to the probability density function (pdf) of $Wx$ and the probability with respect to the Gaussian pdf, i.e. $G(0, \frac{\|x\|^2}{d}I)$. If $p < O(\sqrt{d})$, the error bound goes to zero with $d \to \infty$. By observation 1, therefore, $p$ random projections are asymptotically optimal for such a white data. ∎

*Example 1* : Consider a specific type of white data $x$ that satisfies the source separation generative model $x = Vs$, where $s_k$ are iid with zero-mean and unit-variance and where $V \in \mathbb{R}^{d \times d}$ is orthonormal. Because $\|x\| / \sqrt{d} = \|s\| / \sqrt{d} \to \sqrt{\mathrm{Var}(s_k)} = 1$ with $d \to \infty$, the observation 2 applies, suggesting that $p$ random projections, for any fixed $p$, are maximally informative as the input dimension goes to infinity.

*Example 2* ("compressed sensing of sparse signals") : Consider $x = Vs$ as in the example 1, but with $s$ being $k$ sparse where the nonzero elements are iid. If $k$ is $O(\log d)$ and $p = ck \log d$, then random projections are optimal because $p < O(\sqrt{d})$.

While random projections are asymptotically Gaussian for white data, we are also interested in evaluating their entropy for large but finite-dimensional data. We now develop an explicit approximation for the expected value of the entropy of $p$ random projections in $d$ dimensions, where both $p$ and $d$ are finite.

Specifically, let us consider $x = Vs$, as in the example 1, where $s_k$ follows a generalized Gaussian (GG) distribution. A random variable $x$ is said to be $GG(x; \alpha, \mu, \sigma^2)$ if its pdf is given by

$$p(x) = \frac{\alpha}{2\sqrt{\beta}\sigma\Gamma\left(\frac{1}{\alpha}\right)} \exp\left(-\left|\frac{x-\mu}{\sqrt{\beta}\sigma}\right|^\alpha\right), \tag{4}$$

for $\sigma, \alpha > 0$ and $\beta = \frac{\Gamma(1/\alpha)}{\Gamma(3/\alpha)}$, where $\mu$ is the mean of the distribution, $\sigma$ is the standard deviation, and $\alpha$ is known as the shape parameter. We will simply denote the distribution by $GG(\alpha)$ wherever $\mu$ and $\sigma$ are not specific. The GG includes a number of well-known pdfs as its special cases: $GG(1)$ is a Laplacian exponential distribution, $GG(2)$ corresponds to a Gaussian distribution, whereas in the limiting cases where $\alpha \to 0$, a degenerate distribution in $x = \mu$ is obtained. In general, if $\alpha < 2$, the distribution is sparse, with the degree of the sparsity determined by $\alpha$ (the smaller, the sparser). The shape term of $GG(\alpha)$ is computed to

$$c_\alpha = \frac{1}{2}\ln\left(\frac{4}{\alpha^2}\frac{\Gamma^3\left(\frac{1}{\alpha}\right)}{\Gamma\left(\frac{3}{\alpha}\right)}\right) + \frac{1}{\alpha} \tag{5}$$

in nats, as drawn in Fig 2.

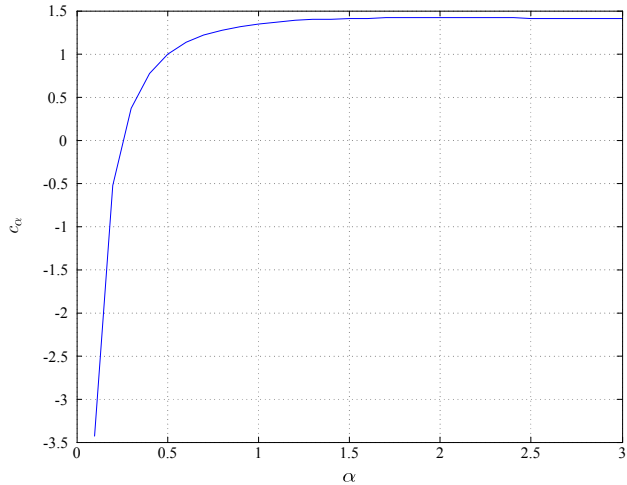

Fig. 2.   Unit-variance entropy (shape term) of $GG(\alpha)$ for various values of $\alpha$.

The net increment of entropy (in its expectation) by adding a new $k$th random projection $y_k$ to preselected $(k-1)$ random projections $\{y_i\}_{i=1}^{k-1}$ is represented by a conditional entropy $E[h(y_k|y_1, \ldots, y_{k-1})]$, which we will call the individual capacity of the random projection and denote by $\nu(k)$.[2] It is easy to see that $\nu(k)$ monotonically decreases along $k$ because, for any $i < j$, $\nu(j) = E[h(y_j|y_1, \ldots, y_{i-1}, \ldots, y_{j-1})]$ $\leq E[h(y_j|y_1, \ldots, y_{i-1})]$ that is equal to $\nu(i)$ due to the symmetry among $y_i$'s. For the white data $x = Vs$, all the random projections are kept uncorrelated with each other as long as they satisfy the orthonormal constraint $WW^T = I$. However, the uncorrelatedness does not lead to independency, which is manifest by the following: From the observation 2 or by the central limit theorem, the first random projection will be maximally informative, so

$$h(y_1) \geq h(s_k), \ \forall k. \tag{6}$$

---

[2] Note that this quantity is *relative* (i.e. meaningful only in comparisonal sense) because it is a kind of differential entropy. Although an individual capacity may take a negative value, including the projection "additionally" to the preselected set of projections is always better than not.

At another extreme, any nondegenerate $d$-dimensional projections should have the same capacity because they are bijective and perfectly describe the original signal $x$, so

$$\sum_k \nu(k) = \sum_k h(s_k) = h(x), \tag{7}$$

which may be understood as *total capacity preservation*. To satisfy (6) and (7) at the same time, the individual capacity of random projections should decrease. Since we consider white data whose the projections are uncorrelated with each other, we imagine the dependency among the projections to be *small* and suggest to ignore high-order multi-information terms.

*Observation 3:* For white data $x = Vs$, where $V \in \mathbb{R}^{d \times d}$ is orthonormal and $s$ has a GG distribution given by (13), the expected value of the entropy of $p$ random projections is

$$E[h(y_1, \ldots, y_p)] \approx pc_2 - \frac{p(p-1)}{d-1}(c_2 - c_\alpha) \tag{8}$$

for large $d$, where $c_\alpha$ denotes the shape term of a $GG(\alpha)$ random variable and $c_2$ is a particular quantity when $\alpha = 2$ (i.e. the shape term of a Gaussian random variable), if we ignore higher-order multi-information terms (than pairwise dependency).

*Proof:* The joint entropy of $k$ random projections $y_1, \ldots, y_k$ can be approximated by

$$h(y_1, \ldots, y_k) \approx \sum_i h(y_i) - \sum_{i<j} I(y_i; y_j) \tag{9}$$

if we neglect higher-order multi-information terms. Again due to the symmetry among $y_i$'s, we may write

$$E[h(y_i)] = h_c, \ \forall i, \qquad E[I(y_i; y_j)] = I_c, \ \forall i \neq j. \tag{10}$$

Inserting (10) into (9), we obtain

$$E[h(y_1, \ldots, y_k)] \approx kh_c - \frac{k(k-1)}{2}I_c. \tag{11}$$

 

If $d$ is sufficiently large, $h_c$ approximates to $c_2$ by the central limit theorem, and $I_c$ is determined to $\frac{2}{d-1}(c_2 - c_\alpha)$ by the total capacity preservation rule. After all, we have (8), deviated from a Gaussian by $\frac{p(p-1)}{d-1}(c_2 - c_\alpha)$. Interestingly, such an order $O(p^2/d)$ appears also in [18] (e.g. see (3)). ■

The observation 3 also tells us that $\nu(k)$ decreases approximately linearly under the assumption that we may ignore the high-order multi-information terms because

$$\nu(k) = E[h(y_1, \ldots, y_k)] - E[h(y_1, \ldots, y_{k-1})] = c_2 - \frac{2(k-1)}{d-1}(c_2 - c_\alpha). \tag{12}$$

Indeed, the property of random projections that makes $\nu(k)$ tilted is attractive for informative sensing because it concentrates large capacity on the first $p$ projections while keeping the remaining uncertainty (i.e. $\sum_{k=p+1}^{d} \nu(k)$) small. Imaginably, the ideal projections would be the one that concentrates "all" capacity on the first few projections (i.e. with capacity falling like a mirrored *step* function). In reality, there might exist no such ideal projections that should form an exact Gaussian as long as they are non-degenerate, yet the *linear* concentration property maintains the random projections still close to Gaussian. In asymptotic case ($d \to \infty$), the linear concentration property meets $\nu(k) \approx c_2$ for any fixed $k$, making $p$ random projections, for any fixed $p$, look like a Gaussian and thus be apparently optimal.

For the white data $x = Vs$, where $p(s)$ is given by

$$p(s) = \prod_{k=1}^{d} GG(s_k; \alpha, 0, 1), \tag{13}$$

for some values of $\alpha$, $h(y_1, \ldots, y_p)$ of random projection and the PCA projection can be computed to

*1) Random:*

$$E[h(y_1, \ldots, y_p)] \approx pc_2 + \frac{p(p-1)}{d-1}(c_2 - c_\alpha). \tag{14}$$

*2) PCA:*

$$h(y_1, \ldots, y_p) = pc_\alpha. \tag{15}$$

For all $\alpha < 2$, random projection performs better than the PCA (or ICA) projection, as shown in Fig. 3, and the performance gap is amplified when $\alpha$ is small (i.e. the distribution is sparse).

### B. Non-white Signals

*Observation 4:* For Gaussian data, the PCA projection is optimal, i.e., maximally informative.

*Proof:* If the PCA projection happens to be Gaussian, both individual terms of the entropy are maximized: the shape term by the Gaussianity and the variance term by the PCA property. Thus, the overall entropy is also maximized. This situation happens for a Gaussian source. Therefore, this provides a simple proof that the optimal projection for Gaussian must be the first $p$ principal components. ∎

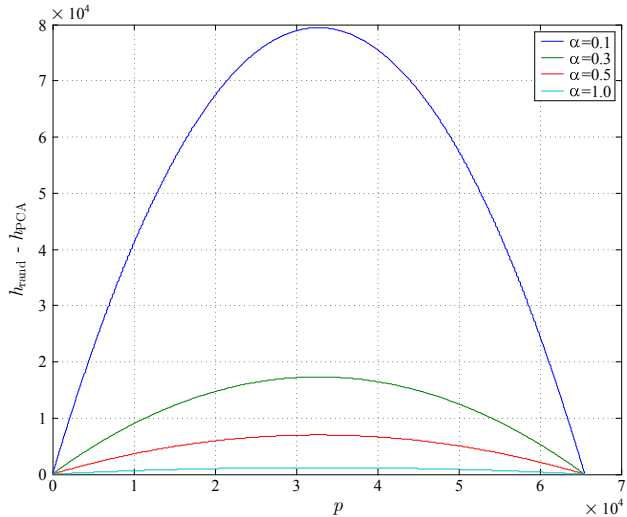

Fig. 3. Relative performance of random projection over PCA projection for a white GG source whose shape parameter is $\alpha$. The original dimension $d$ of the source is $2^{16}$ $(= 65,536)$.

For instance, consider a $d$-dimensional Gaussian distribution whose variance falls off as $1/k^\gamma$ along each axis, $k = 1, \ldots, d$, for a positive value of $\gamma$. If we compute $h(y_1, \ldots, y_p)$ of random projection and the PCA projection, using the entropy decomposition of (2),

*1) Random:*

$$E[h(y_1, \ldots, y_p)] = pc_2 + \frac{1}{2} \ln E \left[ \text{Vol}_p \left( 1, \frac{1}{2^\gamma}, \ldots, \frac{1}{d^\gamma} \right) \right], \tag{16}$$

where $\text{Vol}_p(\lambda_1, \ldots, \lambda_d)$ denotes the volume of a $p$-dimensional hypercube whose edge lengths are randomly (but without repetition) chosen from $\lambda_1, \ldots, \lambda_d$ (see Appendix B for how to compute $\text{Vol}_p(\cdot)$).

*2) PCA:*

$$h(y_1, \ldots, y_p) = pc_2 - \frac{\gamma}{2} \sum_{k=1}^{p} \ln k. \tag{17}$$

For all $\gamma > 0$, the PCA projection performs better than random projection, as shown in Fig. 4, and greater $\gamma$ results in larger performance gap.

Now consider a hybrid (i.e. sparse but non-white) type of the preceding two examples so that

$$p(x) = \prod_{k=1}^{d} GG \left( x_k; \alpha, 0, \frac{1}{k^\gamma} \right). \tag{18}$$

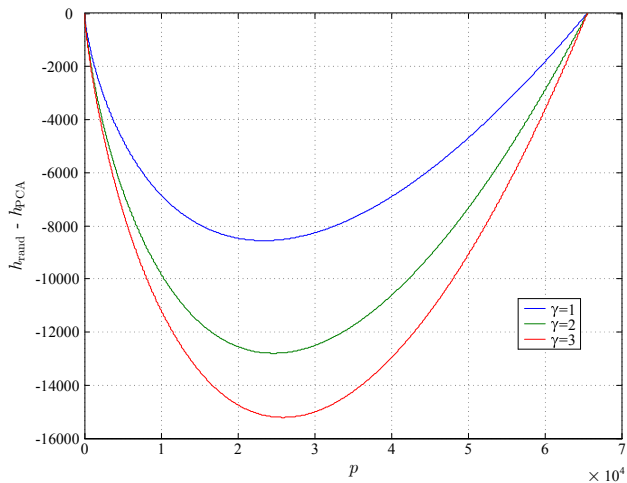

Fig. 4. Relative performance of random projection over PCA projection for a Gaussian source whose variance in each $k$th axis falls off as in $1/k^\gamma$. The original dimension $d$ of the source is $2^{16}$ $(= 65,536)$.

We can simply approximate the relative performance of random projection over the PCA projection by summing two graphs, each from Figs. 3 and 4. As illustrated in Fig. 5, either one may not consistently outperform the other for all range of $p$ (see the cases of $\alpha = 0.3$, $\gamma = 2$ in Fig. 5(a) and $\alpha = 0.5$, $\gamma = 1$ in Fig. 5(b)). However, if $\alpha = 0.5$ and $\gamma = 2$, for example, the PCA projection always does better than random projection.

### C. Noisy Measurement

*Observation 5:* If the noise variance $\sigma^2$ is large, the PCA projections are optimal.

*Proof:* This was proven in [16] and makes sense because major principal components are most durable to noise. Another sketch of the proof can be given as follows: The noisy measurement process $y = Wx + \eta$ can be rewritten as

$$y = W(x + \eta_x) \tag{19}$$

where $\eta_x \sim G(0, \sigma^2 I)$ for an orthonormal matrix $W$. Then, we can pretend as if $x' = x + \eta_x$ were the original source under the noiseless measurement process, for the simple purpose of maximizing $h(y)$. If $\sigma$ is large, $\eta_x$ dominates $x$ in $x' = x + \eta_x$, making the overall distribution Gaussian. Then, from observation 4, the PCA projections are optimal. ∎

Indeed, the rate of convergence (to Gaussian) with increase of $\sigma$ is fast. For scalar random variables $x \sim GG(\alpha, 0, \lambda)$ and $\eta_x \sim G(0, \sigma^2)$, the shape term $c'_\alpha$ of $x' = x + \eta_x$ can be computed by (see

 

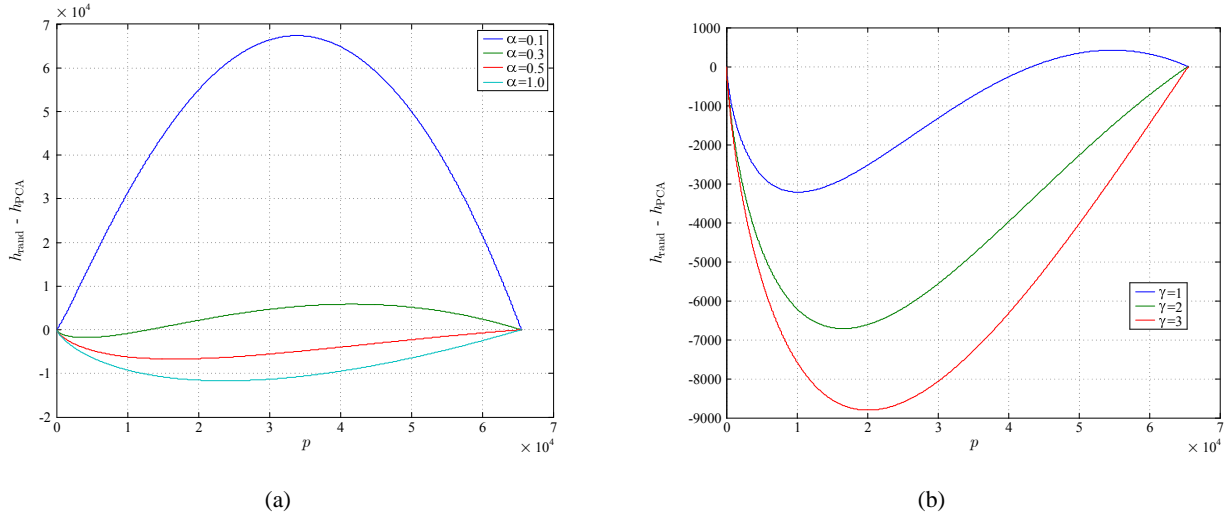

(a)

(b)

Fig. 5. Relative performance of random projection over PCA projection for non-white GG sources. The plots are (a) for various $\alpha$ while $\gamma$ is fixed to 2 and (b) for various $\gamma$ with $\alpha$ fixed to 0.5. The original dimension $d$ of the source is $2^{16}$ ($= 65,536$).

Appendix C)

$$c'_\alpha = h\left(\sqrt{\frac{\lambda/\sigma^2}{1+\lambda/\sigma^2}}\bar{x} + \sqrt{\frac{1}{1+\lambda/\sigma^2}}\bar{\eta}\right),\qquad(20)$$

where $\bar{x} \sim GG(\alpha, 0, 1)$ and $\bar{\eta} \sim G(0,1)$. Therefore, $c'_\alpha$ is a function of $\alpha$ and the signal-to-noise ratio (SNR) $\lambda/\sigma^2$. Unfortunately, we could not find further simplification for (20), but still evaluate it numerically. Fig. 6 illustrates how $c'_\alpha$ changes with respect to the SNR for some values of $\alpha$. As shown in the figure, it grows quite rapidly to $c_2$ (i.e., Gaussian), even with a relatively small amount of noise. In practical applications, measurement noises are often unavoidable and can significantly affect the informativeness of each set of projections.

To summarize, neither random projections nor PCA and ICA are in general the best projections for informative sensing. We showed that for white signals random projection is near-optimal, while for Gaussian signals PCA is optimal. For power-law sparse signals, PCA is better than random unless the signals are extremely sparse. Even for extremely sparse power-law signals, PCA outperforms random with a small amount of noise. Beyond the results, this section also motivates the necessity of a new type of projections, which is universally optimal for informative sensing. Since we actually seek to maximize $h(y)$, the uncertainty of a set of linear projections of data, we call the optimization scheme *uncertain component analysis* (UCA).

## IV. Informative Sensing for Multi-Resolution Signal Models

As we saw in Fig. 5, in general neither random projections nor PCA projections are optimal for non-white signals. In this section, we will consider some practically important class of signals that can be represented by multi-resolution signal models [19]

$$p(x) = \prod_{\ell=0}^{L} \prod_{k \in B_\ell} \frac{1}{\sqrt{\lambda_\ell}} \psi \left( \frac{v_k^T x}{\sqrt{\lambda_\ell}} \right) \tag{21}$$

where $v_k$ and $\lambda_\ell$ denote the independent components of $x$ and the variance of $s_k = v_k^T x$. This representation is based on the source separation generative model $x = Vs$ where $V$ is an orthonormal mixing matrix whose columns consist of the independent components, i.e., $V = [v_1 \ \ldots \ v_d]$. In this model, the independent components are grouped into $(L+1)$ different bands by their resolution and the independent components of the same band (e.g. $B_\ell$) are assumed to be iid, having the same pdf $\psi$ with the same variance $\lambda_\ell$. We will assume, without loss of generality, that $\lambda_0 \geq \lambda_1 \geq \lambda_2 \geq \cdots \geq \lambda_L$. Often, the variance gap between two adjacent bands is known to be quite large, i.e., $\lambda_\ell \gg \lambda_{\ell+1}$, and $\psi$ is modeled by $GG(\alpha)$ for some $\alpha$.

Natural images are a good example of such signals [20]. The independent components $\{v_k\}$ for a natural image are a set of Gabor-like filters in multi-level resolutions [11], with the variance of $v_k^T x$

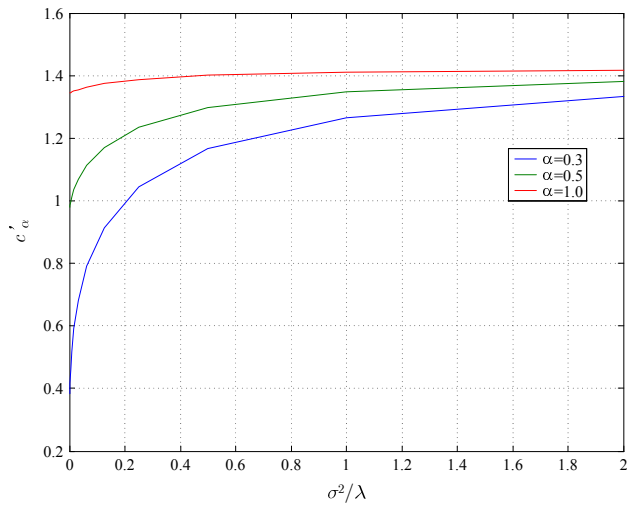

Fig. 6. Variation of the shape term of $GG(\alpha)$ by adding Gaussian noise. Note that the horizontal axis indicates the reciprocal of the SNR, not the SNR itself.

satisfying

$$\text{Var}(v_k^T x) = \lambda_\ell \sim O\left(\frac{1}{4^\ell}\right), \ \forall k \in B_\ell \tag{22}$$

remarkably well, according to the power spectral statistics [21]. The heavy-tailedness in the distribution of $v_k^T x$ is also well known and has been widely modeled by $v_k^T x \sim GG(\alpha)$ with $\alpha \leq 1$ in the literature [22]–[24].

## A. Derivation of UCA Projections

Despite the simplicity of the signal model in (21), it seems infeasible to derive the true optimal projection that maximizes $h(y)$. Here we refrain ourselves from mixing $s_k$'s across bands and seek to find a suboptimal solution among the *bandwise* projections in which $W$ is in the form of

$$W = \begin{bmatrix} W_0 & 0 & \cdots & 0 \\ 0 & W_1 & \cdots & 0 \\ \vdots & \vdots & \ddots & \vdots \\ 0 & 0 & \cdots & W_L \end{bmatrix} V^T. \tag{23}$$

In (23), each submatrix $W_\ell$ is of $p_\ell \times |B_\ell|$, with $p_\ell$ variable while satisfying $\sum_{\ell=0}^{L} p_\ell = p$. If a matrix $W'$ that is not bandwise itself can be made bandwise by rotations, i.e. by premultiplying a unitary matrix $U$, it is congruent to the bandwise matrix $UW'$ because $h(UW'x) = h(W'x)$. In other words, a bandwise matrix $W$, with a particular set of $p_\ell$'s, simulates a bunch of (although not all) matrices $W'$ whose power profile is the same as the distribution of $\{p_\ell\}$, i.e., $\sum_{i=1}^{p} \sum_{j \in B_\ell} [W'V]_{ij}^2 = p_\ell$. Thus, our bandwise restriction in fact includes more projections than explicitly shown in (23).

With this restriction on the projection matrix structure, we have the following observation:

*Observation 6:* For the signal model in (21), if we consider only the bandwise projections in the form of (23) with $p_\ell$'s fixed, random $W_\ell$ are near-optimal as $|B_\ell|$ goes to infinity.

*Proof:* By construction of our model, any bandwise projections taken from different bands are mutually independent, which implies that the optimal set of projections for each band can be found separately. Since each band is white, observation 2 and subsequent arguments can apply here, which gives us bandwise *random* projection. ∎

Now the only remaining job is to determine $p_\ell$. For each band $B_\ell$, the expected value of the entropy behaves like (12) with $d$ substituted by $|B_\ell|$. To illustrate this, Fig. 7 shows the individual capacity $\nu(k)$ of the bandwise random projections, where $\Delta_\ell$ and $\Delta_{\ell+1}$ denote the shape term improvement by randomly mixing the $s_k$'s of the same band, approximating to $c_2 - c_\alpha$, and the slopes are $\frac{2(c_2-c_\alpha)}{|B_\ell|-1}$ and $\frac{2(c_2-c_\alpha)}{|B_{\ell+1}|-1}$, respectively. As shown in the figure, the last few random projections on $B_\ell$ have no much value, in comparison with the first few random projections on $B_{\ell+1}$. Once after evaluating $\nu(k)$ for all $k$, we can simply select the projections associated with $p$ largest values, for the optimal choice.

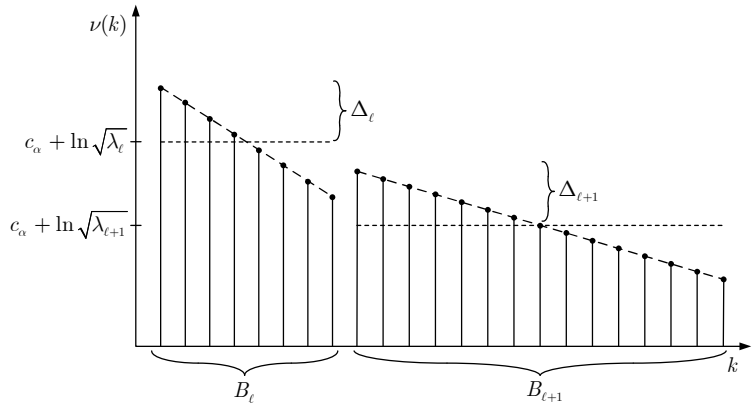

Fig. 7.   Individual capacity diagram for bandwise random projections.

### B. Examples

Considering $256 \times 256$ images ($d = 65,536$), we computed the individual capacity of bandwise random projections, in Fig. 8, for two cases, $\alpha = 0.32$ and $\alpha = 0.49$. For the optimal choice, we should arrange $\nu(k)$ in a non-increasing order and pick the first $p$ projections.

Note that the optimal set of projections varies according to the value of $\alpha$. For $\alpha = 0.49$, the lower frequency bands are far more favored – in other words, truly random projections are avoided – than for $\alpha = 0.32$, due to the relative "denseness" of the source distribution.

### C. Noisy Measurement

Next, we consider noisy measurements on the multi-resolution signal model. If we compute the pdf of $x' = x + \eta_x$ in (19) by convolving $p(x)$ in (21) and $G(\eta_x; 0, \sigma^2 I)$ together, we obtain (see Appendix D)

$$p(x') = \prod_{\ell=0}^{L} \prod_{k \in B_\ell} \frac{1}{\sqrt{\lambda_\ell + \sigma^2}} \psi'_\ell \left( \frac{v_k^T x'}{\sqrt{\lambda_\ell + \sigma^2}} \right) \qquad (24)$$

for some unit-variance pdfs $\psi'_\ell(\cdot)$. This implies that the independent components $\{v_k\}$ are preserved even with the addition of noise $\eta_x$ but the shape and variance of each projection $s'_k = v_k^T x'$ change from those of $s_k$. We see, in (24), the variance increase uniformly by $\sigma^2$. The shape term of $s'_k$ exactly becomes (20) only with $\lambda$ substituted by $\lambda_\ell$ when $k \in B_\ell$.

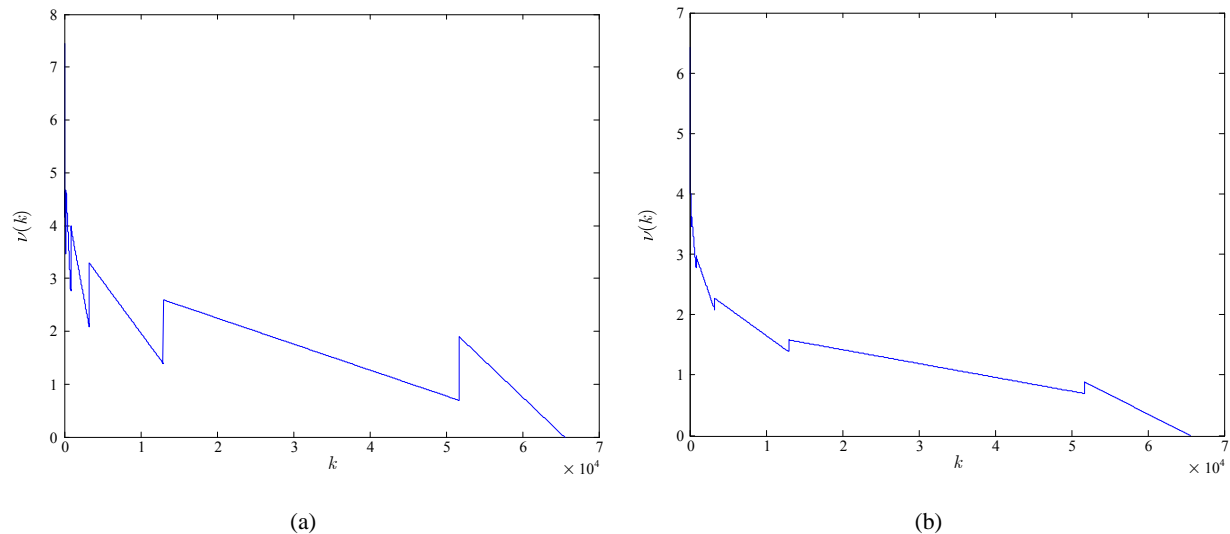

(a)                            (b)

Fig. 8. Individual capacity diagrams of bandwise random projections (a) for $\alpha = 0.32$ and (b) for $\alpha = 0.49$.

The change in variances and shape terms seriously affects the individual capacity diagram. Fig. 9 shows the variation of the individual capacity diagram for the case $\alpha = 0.32$ for various noise levels. As shown in the figure, the overall profile changes drastically even with a relatively small amount of noise. In particular, the slopes in high-frequency bands become flattened, which implies that the random mixing cannot overcome the barrier (variance gap) between the bands, and as a result, low-frequency components are favored. This provides yet another illustration of the observation 5.

## V. EXPERIMENTS: INFORMATIVE SENSING OF NATURAL IMAGES

In this section, we apply the UCA scheme (i.e. bandwise random projections), found in Section IV, to natural images and make comparisons against other kinds of projections (e.g. PCA projection and random projections) in terms of signal reconstruction performance. To implement the proposed UCA scheme, we actually conduct the band decomposition in discrete cosine transfrom (DCT) domain, as illustrated in Fig. 10. instead of explicitly using Gabor-like filters. The DCT kernels are also known to well approximate the principal components of natural images and each kernel in $B_\ell$ represents some harmonic (deterministic) mixing of the independent components that lie in the frequencies between

$$\frac{2^\ell}{4\sqrt{d}} \leq \frac{f}{f_s} < \frac{2^\ell}{2\sqrt{d}} \tag{25}$$

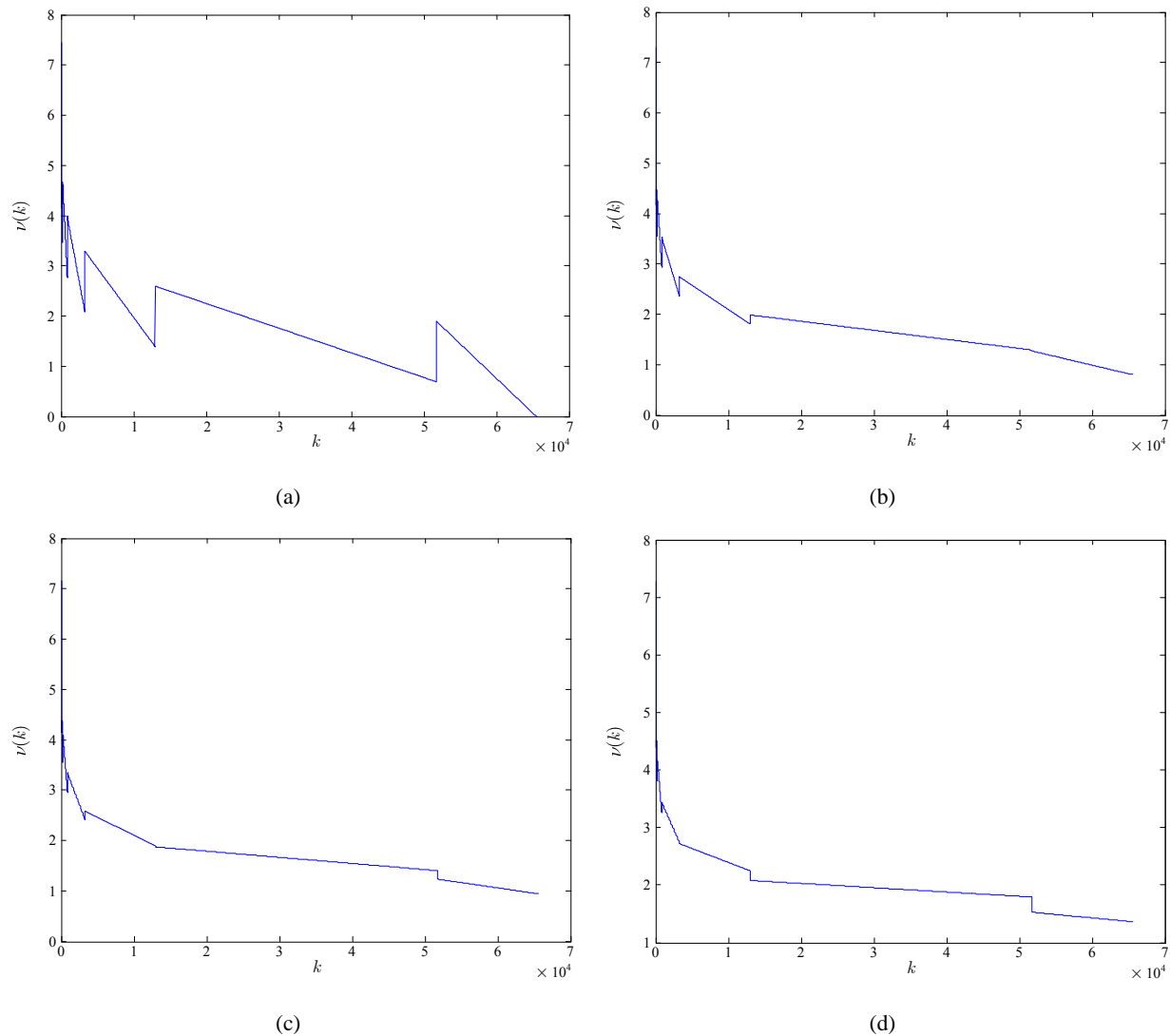

Fig. 9. Individual capacity diagrams for the Cameraman image for various noise level. (a) $\sigma = 0$ (no noise), (b) $\sigma = 5$, (c) $\sigma = 10$, (d) $\sigma = 20$. Each pixel can have a value from 0 to 255. The noise level may be understood in comparision with the maximum pixel value.

where $f = \sqrt{f_x^2 + f_y^2}$ and $f_s$ denotes the sampling frequency in both directions. Then, the band selection in DCT domain can effectively sift the independent components of the same resolution. In fact, the independent components are over-complete,[3] but we assume as if there were only a complete set of independent components orthogonal to each other.[4] Then, random mixing of DCT coefficients on a specific band is treated equivalent to random mixing of the independent components in that band.

---

[3]In a specific band (resolution), each independent component corresponds to a local edge at a particular location and angle.

[4]In other view, we are approximating the smooth power spectrum falling off as $1/f^2$ by a bandwise-flat one.

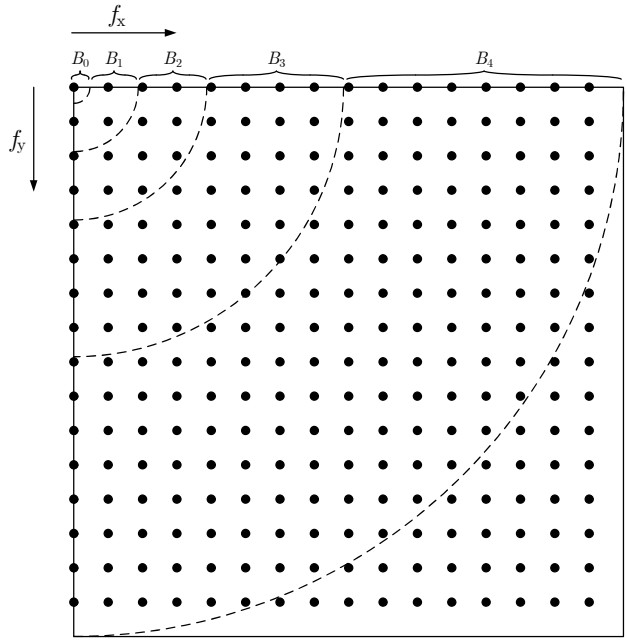

Fig. 10. Illustration of band decomposition in DCT domain.

To carry out random mixing, specifically we use a set of noiselets [25], binary-valued random matrix, for the efficient computer simulation. In doing so, for further ease of simulation, we make a slight modification to the UCA scheme. Given the total number of projections $p$, we determine the number of projections $p_\ell$ for each band $B_\ell$ utilizing the capacity diagram. In practice, we take all $|B_\ell|$ projections if $p_\ell > 0.9 \, |B_\ell|$, while taking none if $p_\ell < 0.1 \, |B_\ell|$, which removes the necessity of random mixing in both cases. Then, we take the remaining number of random projections across all the other bands at a time.

The signal reconstruction experiments have been built on the basis of Romberg's implementation [26]. To reconstruct an image from measurements, we minimized the total variation (TV) of $\widehat{x}$, subject to $y = W\widehat{x}$, defined by

$$\|\widehat{x}\|_{\mathrm{TV}} = \sum_{i,j} \left| \nabla \widehat{X}_{ij} \right|, \tag{26}$$

where $\widehat{X}$ is the matrix representation of $\widehat{x}$. The TV minimization is known to perform better than the $L_1$-norm minimization on the sparse basis (e.g. wavelets), avoiding high-frequency artifacts [26], [27].

The experimental results obtained for a couple of $256 \times 256$ images, *Cameraman* and *Einstein*, are shown in Fig. 11. We compared the performance, in terms of peak-signal-to-noise ratio (PSNR), among five different schemes: linear reconstruction using low-pass DCT coefficients in zig-zag order (blue), nonlinear reconstruction using the same set of DCT coefficients (green), Romberg's method [26] which takes the first 1,000 DCT coefficients, also in zig-zag order, and switches to random projections (red), pure random projections (cyan), and bandwise random projections, which are the uncertain components we found for natural images (magenta). Except for the first one, TV minimization has been commonly used to recover images from each set of measurements.

Unsurprisingly, in every case, the green curve (DCT with TV minimization) is above the blue (DCT with linear reconstruction), and the red curve (1k DCT + random projections) is above the cyan (pure random projections). However, if we look at the green (DCT with TV minimization) and the red (1k DCT + random projections), their relative performance changes completely, depending on the source image. Indeed, the two images turn out to have quite different characteristics in terms of their sparsity. The GG shape parameter was estimated, from their Haar wavelet coefficients, to $\alpha \approx 0.32$ for the Cameraman image and to $\alpha \approx 0.49$ for the Einstein image, with their capacity diagrams corresponding to Fig. 8(a) and 8(b), respectively.

The Cameraman image is quite sparse. A moderate number of random projections are capable of evenly grabbing image contents from all spatial frequencies. Meanwhile, the DCT projection uses up all available sensors only for the low-frequency contents, which could be captured with even fewer sensors, while missing nearly all high-frequency details. On the other hand, the Einstein image is not that sparse. As suggested by Fig. 8(b), we must deploy almost all sensors for low-frequency bands. Otherwise, even low-resolution version of the image cannot be recovered faithfully. Indeed, the DCT projection proves almost optimal for the Einstein image.

The UCA projections (magenta) outperform Romberg's method (red) as well as the DCT projections (green). In principle, the UCA projections are expected to achieve at least the upper bound of the two. On occasion, however, for the Einstein image, the DCT projection was marginally better than the UCA projection. This is because the DCT projection is nearly optimal for the Einstein image and the UCA projection, based on (21), may suffer from inaccurate modeling artifacts.

Fig. 12 shows the reconstruction results of 5,000 measurements (7.6% of the original dimension) on the Cameraman image, which portray the behavioural characteristics of each scheme. First, the image linearly reconstructed from DCT projections is blurred and also ringing. Such artifacts can be removed by employing a nonlinear reconstruction (i.e. TV minimization). However, the measurements were still concentrated on low-frequencies, so the image almost loses significant mid/high-frequency contents. In contrast, Romberg's method pursues high-frequency contents too hastily despite the seriously limited number of measurements. Indeed, it somewhat succeeds in recovering high-frequency details, but with much sacrifice of the low/mid-frequency contents which is more important. Last, the UCA projection gives up the high-frequency contents but preserves low/mid-frequency contents quite faithfully.

We did similar experiments also in noisy settings. In this case, we concatenated a denoising module based on field-of-experts image prior model [24] after the TV minimization, for all nonlinear reconstruction schemes. Fig. 13 compares the performance of the five compressed sensing schemes for $p = 21,000$ at various noise levels. Note that, for the Cameraman image, having started worst among nonlinearly recovered schemes, the DCT projection (green) catches up and even exceeds all the others as $\sigma$ increases, while Romberg's method (red) as well as random projections (cyan) degrade fast. For the Einstein image, the DCT projection is persistently better than Romberg's method and random projections, and more remarkably, the degradation proceeds slowest. The UCA projection (magenta) finds the best set of projections throughout most range of noise levels, converging to the DCT projection as $\sigma$ increases. In the low SNR regime, the UCA projection worked slightly worse than the DCT projection, perhaps due to the inaccurate modeling artifacts again.

Note that the UCA scheme uses different sets of projections depending on the sparsity of the source image and also on the noise level. In case that the sparsity of the source image is unknown, we might have to use a value learnt in advance, from a large collection of natural images. Then, we can achieve near-optimal performance in overall sense, but not in every individual case. In certain applications, it may be allowed to sense a few hundred Haar wavelet coefficients so that we can estimate the sparsity before we do tens of thousands of projections.

## VI. CONCLUSION

Suppose we are allowed to take a small number of linear projections of signals in a dataset, and then use the projections plus our knowledge of the dataset to reconstruct the signals. What are the best projections to use? We have shown that these projections are not necessarily the principal components nor the independent components of the data nor random projections, but rather a new set of projections which we call uncertain components. We formalized this notion, informative sensing, by maximizing the mutual information between the signals and their projections.

Then, we presented some analytical results which help us to understand the desirable behaviors for the informative sensing. For white data, the most informative projections are those that are as Gaussian as possible, in favor of random projections, while PCA or ICA can significantly outperform random projections for highly non-white data or in low SNR regime. In particular, for natural images, we showed that more sensors should be reserved for low-frequency contents than used for high-frequency contents but not too many, which makes bandwise random projections most informative.

## Appendix

### A. Derivation of $h(Wx)$ for the Objective Function in Noiseless Settings

Consider the subspace $\mathcal{W}_\perp$ not spanned by the row vectors of $W$, which represents unmeasured dimension of $x$, and define $W_\perp$ so that its row vectors be orthonormal bases for $\mathcal{W}_\perp$. If we define

$$\begin{bmatrix} y \\ z \end{bmatrix} \stackrel{\text{def}}{=} \begin{bmatrix} Wx \\ W_\perp x \end{bmatrix} = \begin{bmatrix} W \\ W_\perp \end{bmatrix} x = Ux, \tag{27}$$

the pair of $(y, z)$ corresponds to $x$ in *rotated* bases because of the unitarity of $U$. Since we exactly measure $y$, some partial coordinates of $x$, the remaining job is to infer $z$ using $y$ at hand. In doing so, to reduce the uncertainty about $z$ as much as possible, we seek to minimize $h(z|y)$, and from the fact that $h(x) = h(Ux) = h(y, z) = h(y) + h(z|y)$ and $h(x)$ is fixed, minimizing $h(z|y)$ is just equivalent to maximizing $h(y)$, for the noiseless condition.

### B. Subvolume Expectation

Let $\Lambda_m = \{\lambda_1, \lambda_2, \ldots, \lambda_m\}$ and define $S_p(\Lambda_m)$ as the sum of all the products made up of $p$ elements in $\Lambda_m$. Then,

$$E[\mathrm{Vol}_p(\lambda_1, \ldots, \lambda_d)] = \frac{S_p(\Lambda_d)}{\binom{d}{p}}. \tag{28}$$

Note that $S_p(\Lambda_m)$ can be specified recursively by

$$S_p(\Lambda_m) = \sum_{j=p}^{m} S_{p-1}(\Lambda_{j-1})\lambda_j, \quad p, m = 1, 2, \ldots, d, \tag{29}$$

with $S_0(\cdot) = 1$, which enables us to efficiently compute $E[\mathrm{Vol}_p(\lambda_1, \ldots, \lambda_d)]$ by dynamic programming.

### C. Derivation of (20)

Because $x' = x + \eta_x = \sqrt{\lambda}\bar{x} + \sigma\bar{\eta}$, where $\bar{x} \sim GG(\alpha, 0, 1)$ and $\bar{\eta} \sim G(0, 1)$. the variance of $x'$ is $\lambda + \sigma^2$, and by definition of the shape term,

$$c'_\alpha = h\left(\frac{x'}{\sqrt{\lambda + \sigma^2}}\right) = h\left(\frac{\sqrt{\lambda}\bar{x} + \sigma\bar{\eta}}{\sqrt{\lambda + \sigma^2}}\right) \tag{30}$$

which can be simply arranged into (20).

*D. Derivation of (24)*

The pdf of $x' = x + \eta_x$ can be computed by convolving $p(x)$ and $p(\eta_x)$ because $\eta_x$ is independent of $x$. Changing the variable $x$ by $s = V^T x$, where $V = [v_1 \ \ldots \ v_d]$, and exploiting $p(\eta_x) = G(\eta_x; 0, \sigma^2 I) = \prod_{\ell=0}^{L} \prod_{k \in B_\ell} G(v_k^T \eta_x; 0, \sigma^2)$,

$$
\begin{aligned}
p(x') &= \int p(x) G(x' - x; 0, \sigma^2 I) dx \\
&= \int p(Vs) G(x' - Vs; 0, \sigma^2 I) ds \\
&= \int \prod_{\ell,k} \frac{1}{\sqrt{\lambda_\ell}} \psi\left(\frac{s_k}{\sqrt{\lambda_\ell}}\right) \prod_{\ell,k} G(v_k^T x' - s_k; 0, \sigma^2) ds \\
&= \prod_{\ell,k} \int \frac{1}{\sqrt{\lambda_\ell}} \psi\left(\frac{s_k}{\sqrt{\lambda_\ell}}\right) G(v_k^T x' - s_k; 0, \sigma^2) ds_k \\
&= \prod_{\ell,k} \phi'_\ell(v_k^T x'),
\end{aligned}
\tag{31}
$$

where

$$
\phi'_\ell(\tau) = \int \frac{1}{\sqrt{\lambda_\ell}} \psi\left(\frac{\omega}{\sqrt{\lambda_\ell}}\right) G(\tau - \omega; 0, \sigma^2) d\omega.
\tag{32}
$$

The variance $(\sigma'_k)^2$ of $s'_k = v_k^T x'$ can be easily calculated by

$$
(\sigma'_k)^2 = \text{Var}\left(v_k^T x + v_k^T \eta_x\right) = \lambda_\ell + \sigma^2, \qquad \forall k \in B_\ell.
\tag{33}
$$

Finally, defining $\psi'_\ell(\tau)$ as $\psi'_\ell(\tau) = \sqrt{\lambda_\ell + \sigma^2} \phi'_\ell(\sqrt{\lambda_\ell + \sigma^2} \tau)$, we obtain (24).

## REFERENCES

[1] D. L. Donoho, "Compressed sensing," *IEEE Trans. Inf. Theory*, vol. 52, no. 4, pp. 1289–1306, Apr. 2006.

[2] E. J. Candès and T. Tao, "Near-optimal signal recovery from random projections: Universal encoding strategies?" *IEEE Trans. Inf. Theory*, vol. 52, no. 12, pp. 5406–5425, Dec. 2006.

[3] M. F. Duarte, M. B. Wakin, D. Baron, and R. G. Baraniuk, "Universal distributed sensing via random projections," in *Proc. of International Conference on Information processing in Sensor Networks*, Nashville, TN, Apr. 2006, pp. 177–185.

[4] J. Haupt and R. Nowak, "Signal reconstruction from noisy random projections," *IEEE Trans. Inf. Theory*, vol. 52, no. 9, pp. 4036–4048, Sept. 2006.

[5] M. J. Wainwright, "Sharp thresholds for high-dimensional and noisy recovery of sparsity," in *In Proc. Allerton Conference on Communication, Control and Computing*, 2006.

[6] R. M. Castro, J. Haupt, R. Nowak, and G. M. Raz, "Finding needles in noisy haystacks," in *Proc. IEEE Int. Conf. on Acoustics, Speech and Signal Processing*, Mar. 2008, pp. 5133–5136.

[7] M. Duarte, M. Davenport, D. Takhar, J. Laska, T. Sun, K. Kelly, and R. Baraniuk, "Single-pixel imaging via compressive sampling," *IEEE Signal Process. Mag.*, vol. 25, no. 2, pp. 83–91, Mar. 2008.

[8] M. Lustig, J. M. Santos, D. Donoho, and J. M. Pauly, "k-t sparse: High frame rate dynamic mri exploiting spatio-temporal sparsity," in *Proc. Annual Meeting of ISMRM*, 2006.

[9] M. W. Seeger, H. Nickisch, R. Pohmann, and B. Schölkopf, "Bayesian experimental design of magnetic resonance imaging sequences," in *Advances in Neural Information Processing Systems*, 2008.

[10] T. Lin and F. J. Herrmann, "Compressed wavefield extrapolation," *Geophysics*, vol. 72, no. 5, pp. SM77–SM93, Sept./Oct. 2007.

[11] A. J. Bell and T. J. Sejnowski, "Edges are the independent components of natural scenes," in *Advances in Neural Information Processing Systems*, vol. 9, 1997, pp. 831–837.

[12] R. Linsker, "An application of the principle of maximum information preservation to linear systems," in *Advances in Neural Information Processing Systems*, vol. 1, 1989, pp. 186–194.

[13] F. Attneave, "Informational aspects of visual perception," *Psych. Rev.*, vol. 61, pp. 183–193, 1954.

[14] H. B. Barlow, "Possible principles underlying the transformation of sensory messages," in *Sensory Communications*, W. A. Rosenblith, Ed. Cambridge, MA: MIT Press, 1961, pp. 217–234.

[15] J. J. Atick, "Could information theory provide an ecological theory of sensory processing?" *Network: Comput. Neural Syst.*, vol. 3, pp. 213–251, 1992.

[16] Y. Weiss, H. S. Chang, and W. T. Freeman, "Learning compressed sensing," in *Proc. Allerton Conf. on Communication, Control, and Computing*, Sept. 2007.

[17] M. W. Seeger and H. Nickisch, "Compressed sensing and Bayesian experimental design," in *Proc. Int. Conf. on Machine Learning*, June 2008, pp. 912–919.

[18] S. Dasgupta, D. Hsu, and N. Verma, "A concentration theorem for projections," in *Proc. of 22nd Conf. on Uncertainty in Artificial Intelligence*, July 2006.

[19] S. G. Mallat, "A theory for multiresolution signal decomposition: The wavelet representation," *IEEE Trans. Pattern Anal. Machine Intell.*, vol. 11, no. 7, pp. 674–693, July 1989.

[20] M. Bethge, "Factorial coding of natural images: How effective are linear models in removing higher-order dependencies?" *J. Opt. Soc. Am. A*, vol. 23, no. 6, pp. 1253–1268, June 2006.

[21] A. van der Schaaf and J. van Hateren, "Modelling the power spectra of natural images: statistics and information," *Vision Research*, vol. 36, no. 17, pp. 2759–2770, 1996.

[22] B. A. Olshausen and D. J. Field, "Emergence of simple-cell receptive field properties by learning a sparse code for natural images." *Nature*, vol. 381, pp. 607–609, June 1996.

[23] E. P. Simoncelli, "Statistical models for images: Compression, restoration and synthesis," in *Proc. Asilomar Conf. on Signals, Systems and Computers*, Nov. 1997, pp. 673–678.

[24] Y. Weiss and W. T. Freeman, "What makes a good model of natural images?" in *Proc. IEEE Conf. on Computer Vision and Pattern Recognition*, Minneapolis, MN, June 2007.

[25] E. J. Candès and J. Romberg, "Sparsity and incoherence in compressive sampling," *Inverse Prob.*, vol. 23, no. 3, pp. 969–986, June 2007.

[26] J. Romberg, "Imaging via compressive sampling," *IEEE Signal Process. Mag.*, vol. 25, no. 2, pp. 14–20, Mar. 2008.

[27] R. Berinde and P. Indyk, "Sparse recovery using sparse random matrices," MIT, Tech. Rep., 2008.

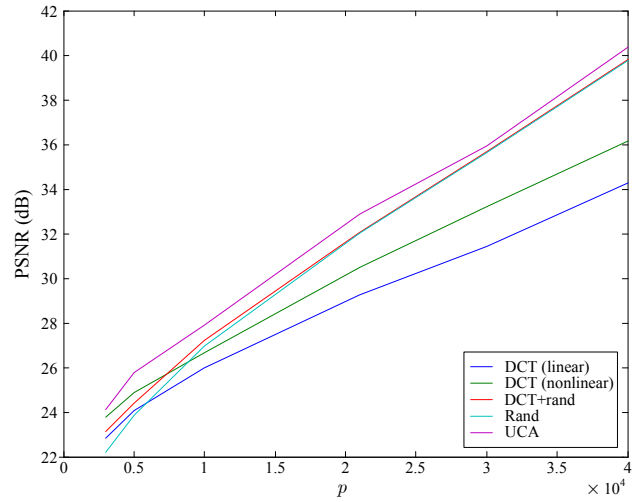

(a)

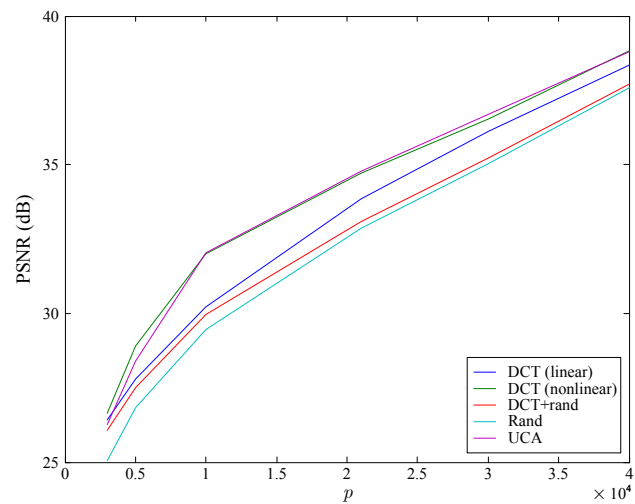

(b)

Fig. 11. Experimental results on two images: (a) Cameraman and (b) Einstein. Compared schemes are DCT with linear reconstruction (blue), DCT with TV minimization (green), 1k DCT + random (red), pure random (cyan), and UCA projection (magenta).

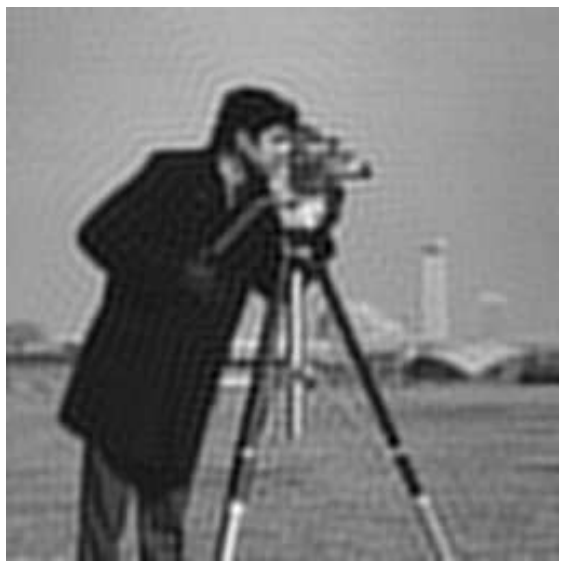

(a)

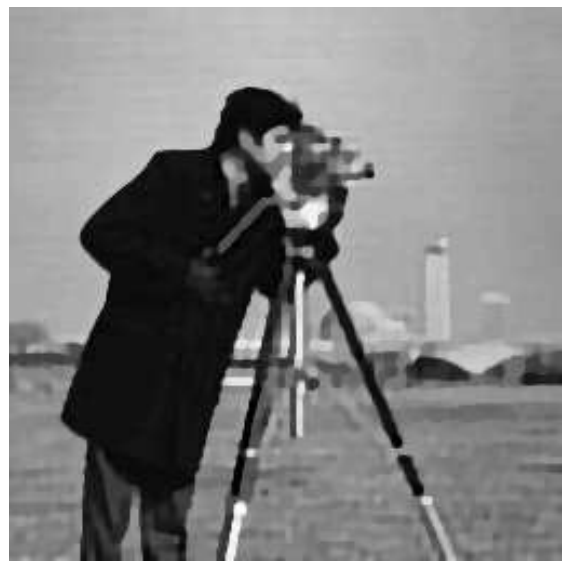

(b)

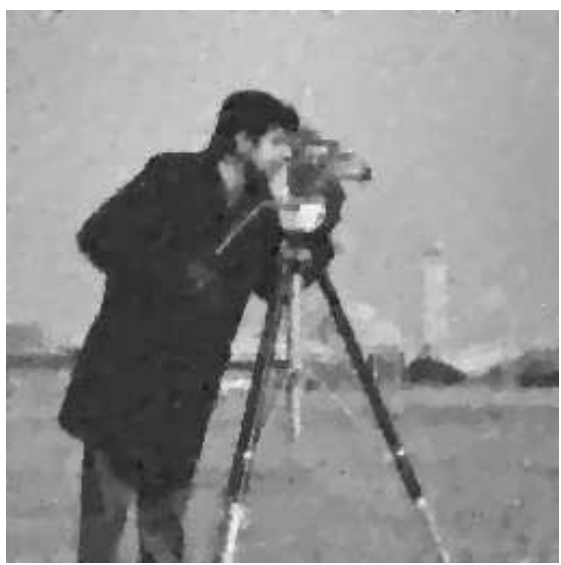

(c)

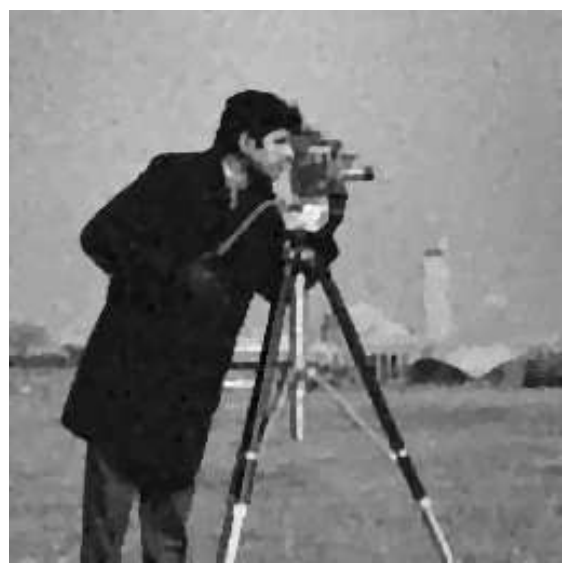

(d)

Fig. 12. Cameraman image reconstruction using 5,000 measurements. (a) DCT with linear reconstruction (24.06dB), (b) DCT with TV minimization (24.88dB), (c) 1k DCT + random (24.41dB), (d) UCA (25.78dB).

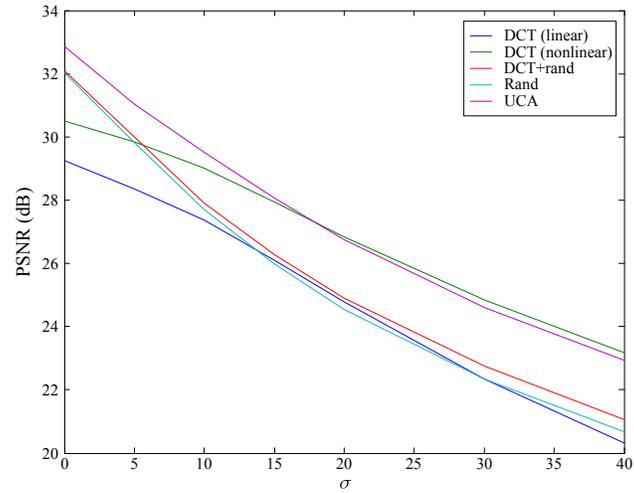

(a)

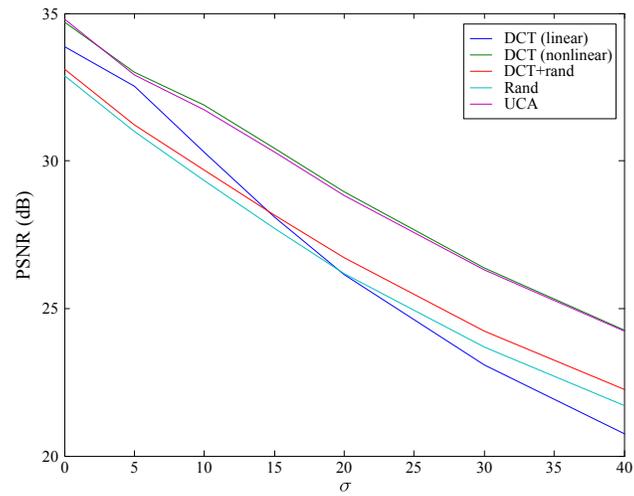

(b)

Fig. 13. Experimental results, in noisy conditions, on two images: (a) Cameraman and (b) Einstein. Compared schemes are DCT with linear reconstruction (blue), DCT with TV minimization (green), 1k DCT + random (red), pure random (cyan), and UCA projection (magenta). The number of measurements are set to $p = 21,000$.